\documentclass{article}

\usepackage[main, final]{neurips_2026}
\usepackage{flafter}
\usepackage{placeins}
\usepackage{float}
\usepackage{multirow}
\usepackage[utf8]{inputenc}
\usepackage[T1]{fontenc}
\usepackage{hyperref}
\usepackage{authblk}
\usepackage{url}
\usepackage{booktabs}
\usepackage{tabularx}
\usepackage{amsfonts}
\usepackage{nicefrac}
\usepackage{microtype}
\usepackage{xcolor}
\usepackage{graphicx}
\usepackage{amsmath}
\usepackage{amssymb} 

\graphicspath{{figures/}}

\title{Qwen-Audio-3.0-Gen-Preview Technical Report}

\author{%
  \parbox{0.78\textwidth}{%
    \centering
    \textit{(Equal contribution; alphabetical by family name.)}\\
    Junyu Dai, Xiaoyue Duan, Xinyue Fan, Yihan Feng,
    Jingbei Li, Xiangang Li, Yunjia Li, Lejun Min, Yufei Shi,
    Xingchen Song, Yiran Wang, Cheng Wen, Menglin Wu,
    Bajian Xiang, Huaicheng Zhang, Han Zhao, Ruichen Zheng\par
  }%
}
\affil{%
  \makebox[\textwidth][c]{%
    \parbox{0.72\textwidth}{%
      \centering\normalsize\bfseries
      \raisebox{-0.20em}{\includegraphics[height=1.05em]{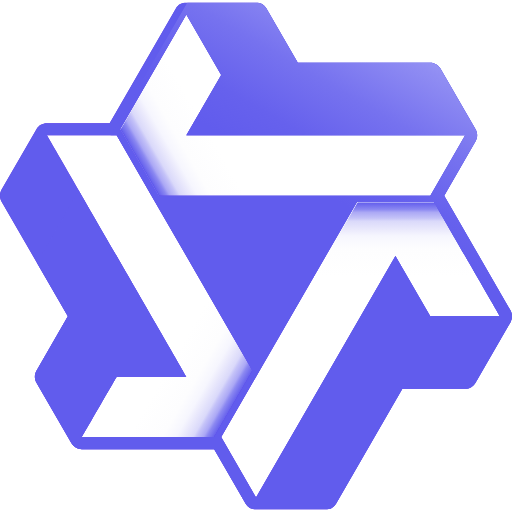}}%
      \enspace Alibaba Token Foundry\par
    }%
  }%
}
\date{}

\begin{document}

\maketitle

\begin{abstract}
Existing single-domain and multi-task audio systems remain limited in directly
organizing heterogeneous audio components, ambience, and multiple roles into
long-form temporal scenes. We present
\textbf{Qwen-Audio-3.0-Gen-Preview}, a unified non-autoregressive
framework that uses a Diffusion Transformer (DiT) and a shared variational
autoencoder (VAE) to generate the complete mixed waveform. Prompt enhancement
converts free-form requests into structured temporal records that are rendered
as textual conditions, while a two-stage data curriculum and semantic
conditional views train the proposed model to use these conditions across
standalone and mixed-scene audio. A shared continuous VAE compresses 48\,kHz
stereo waveforms into 25\,Hz latent sequences and incorporates semantic
supervision, providing one representation for heterogeneous audio. On the
public reference-conditioned benchmark, speaker similarity is the proposed
model's clearest strength across all three subsets. Across the multi-speaker and
rich-timeline benchmarks, its clearest comparative strengths are cross-turn
consistency in both languages and temporal localization, respectively. On
AudioCaps, its advantages are concentrated in evaluations using large
audio-language models and AudioBox.
These results demonstrate the potential of unified generation for temporally
structured audio without task-specific branches.

\end{abstract}

\setcounter{section}{0}
\section{Introduction}

Existing audio generation models can already produce high-quality linguistic
content~\cite{chen2025f5,le2023voicebox,wang2025maskgct,shen2024naturalspeech,chen2025neural},
environmental events~\cite{liu2023audioldm,liu2024audioldm,ghosal2023text},
and long-range structured audio~\cite{copet2023simple,agostinelli2023musiclm}
in separate settings. However, real-world content creation often requires more
than a single type of sound. A scene in a film, game, or podcast may contain
character dialogue, background ambience, action-related events, and structured
background audio at the same time. Producing such a scene typically requires
separate models for different audio components, followed by manual editing to
place the generated clips on a timeline, align them, adjust their relative
levels, and mix them into a final track. This multi-stage process is
time-consuming and makes it difficult to maintain both overall coherence and
precise control over individual audio components.

Recent work on unified audio generation has started to support heterogeneous
audio generation within a single framework~\cite{yang2023uniaudio,vyas2023audiobox}.
These models greatly expand the range of tasks that one model can handle.
However, supporting multiple generation tasks is not the same as generating a
unified audio scene. Existing models often generate audio components in response
to separate instructions, but pay less attention to how these components should
appear together, interact, and develop over time within one continuous track.

The key challenge of complex audio scene generation is not simply to expand the
range of sounds a model can generate, but to organize heterogeneous sounds into
a coherent scene. The model must control which sounds occur, when they start and
end, how they overlap and follow one another, and how their relative levels are
balanced. It must also coordinate characters, dialogue, ambience, foreground
events, and long-range background structure within a shared temporal
organization. For long-form narratives, the model must further preserve
character voices across dialogue turns and maintain continuity in the acoustic
environment over time.


In this report, we present \textbf{Qwen-Audio-3.0-Gen-Preview}, a unified
non-autoregressive (NAR) framework that uses a Diffusion Transformer (DiT)
operating in the continuous latent space of a shared variational autoencoder
(VAE). Through a single generation path, the proposed model supports
heterogeneous standalone audio, multi-speaker dialogue, and temporally
organized mixed scenes. Free-form user requests are converted into structured
temporal conditions, while a two-stage training curriculum first establishes
broad audio generation and text--audio correspondence and then teaches the
proposed model to organize heterogeneous sounds into long-form scenes. To
evaluate capabilities not well captured by existing benchmarks, we further
introduce two in-house benchmarks for reference-conditioned multi-speaker audio
and rich-timeline audio generation. On the public reference-conditioned
benchmark, the proposed model achieves the highest speaker similarity among the
evaluated systems; on the two in-house benchmarks, its clearest comparative
strengths are cross-turn speaker consistency and temporal localization. On
AudioCaps, its clearest advantages occur under large audio-language model and
AudioBox assessments.
Together, these results show that a unified model can retain strong audio
generation capabilities while extending generation toward long-form,
multi-character, and temporally structured scenes.

\setcounter{section}{1}
\section{Related Work}

\subsection{From Domain-Specific Generation to Unified Audio Generation}
Audio generation has traditionally been divided into domain-specific tasks.
Systems for linguistic content focus on intelligibility, naturalness, speaker
similarity, and expressive control
\cite{chen2025f5,wang2025maskgct,le2023voicebox,shen2024naturalspeech,chen2025neural}.
Recent systems have further extended this direction toward long-form and
multi-speaker generation. VibeVoice supports long-form multi-speaker
conversational synthesis through an ultra-low-frame-rate continuous tokenizer
and next-token diffusion \cite{peng2026vibevoice}, while Any2Speech uses
hierarchical annotations to model scene context, speaker profiles,
utterance-level expression, and pronunciation \cite{song2026borderless}.
Other domain-specific systems generate environmental sounds, acoustic events,
or long-range structured audio from natural-language descriptions and related
controls
\cite{liu2023audioldm,liu2024audioldm,ghosal2023text,copet2023simple,agostinelli2023musiclm}.
Although these models achieve high generation quality on their respective
tasks, they often rely on different audio representations, model architectures,
and conditioning interfaces. When an application requires several audio
components at the same time, separate systems still need to be combined.

Unified audio generation aims to support multiple audio domains within shared
representations or a shared model. UniAudio uses a unified discrete audio
representation for a wide range of generation tasks \cite{yang2023uniaudio},
while Audiobox applies flow matching to both linguistic and general audio
generation \cite{vyas2023audiobox}. AudioX \cite{tian2025audiox},
UniFlow-Audio \cite{xu2025uniflow}, UniSonate \cite{qiang2026unisonate} and
Audio-Omni \cite{tian2026audio} further expand the supported input modalities,
generation tasks, understanding capabilities, and editing functions. UNISON
also supports dialogue-and-sound generation, speaker cloning, temporal
composition, and scene-level editing \cite{li2026unison}. These models show
that different audio domains can share representations and model parameters.
However, their main focus is still to expand the range of tasks covered by a
single model. A model may generate different audio components under different
instructions, but may not organize them as connected parts of the same scene.

\subsection{Complex Audio Scene Generation}
Complex audio scene generation focuses on producing a complete mixed track that
contains heterogeneous audio components, persistent ambience, and temporal
structure. Existing methods mainly follow two directions: external coordination
and direct mixed-track generation with a single model.

External coordination methods first convert a user description into a script or
timeline, then call several specialized models to generate different audio
components, and finally combine them through post-processing. WavJourney uses
structured scripts to organize audio generation~\cite{liu2025wavjourney}.
Audio-Oscar further introduces multiple agents for sound design, timeline
planning, model selection, generation, and
post-production~\cite{duan2026audio}. These methods are modular and allow
direct editing of scripts and timelines. However, because different components
are generated independently, overall consistency and cross-component
interaction depend heavily on external mixing. Their multi-stage pipelines also
increase system complexity and inference cost.

Another direction is to generate the final mixed track directly with a single
model. Bagpiper can generate heterogeneous audio mixtures using an
autoregressive audio foundation model \cite{tian2026bagpiper}. Foley-Omni
focuses on generating complete soundtracks conditioned on video
\cite{tao2026foley}. Dasheng AudioGen uses structured scene descriptions to
represent diverse audio components and acoustic environments, and directly
generates the mixture in a shared continuous semantic-acoustic latent space
\cite{mei2026dashengaudiogen}. By modeling all components in one generation
process, these methods can learn interactions between different sounds and
avoid track-by-track generation and assembly. However, existing single-model
approaches are still mainly designed for relatively short clips or simple
combinations of sounds. Long-form scenes require stronger support for
consistent character voices, multi-turn dialogue, and coordinated temporal
structures across dialogue, ambience, foreground events, and long-range
background structure.

\subsection{Temporal Control and Compositional Instruction Following}

To improve controllability in multi-event audio generation, several studies have started to model the timing of sound events. Make-An-Audio 2 uses a large language model to convert natural-language descriptions into structured event sequences and supports variable-length generation \cite{huang2023make}. PicoAudio and PicoAudio2 use temporally aligned supervision to control when events occur and how often they appear \cite{xie2025picoaudio,zheng2026picoaudio2}. ControlAudio combines text, timing, and phoneme conditions to improve both temporal accuracy and speech intelligibility \cite{jiang2026controlaudio}.

Complementary efforts have focused on data construction, feedback-based
training, and evaluation for temporal instruction following.
AudioTime~\cite{xie2025audiotime} and T2A-Feedback~\cite{wang2025t2a} evaluate
event order, duration, frequency, temporal relations, and content completeness
in multi-event audio generation. Together, these efforts broaden temporal
instruction following from isolated event timing to compositional constraints
over multiple events.

However, existing work on temporal control mainly focuses on general sound
events and whether they occur at the specified time. Complex audio scenes
involve a broader problem. The model must also handle character identity,
multi-turn dialogue, persistent ambience, and long-range relations among
different components.

Rather than introducing another domain-specific generator or simply expanding
the task coverage of a unified model, our work focuses on scene-level audio
organization. Heterogeneous audio components are modeled as interacting parts
of the same scene, together with their temporal relations, and jointly rendered
into a complete mixed track in a shared continuous latent space.

\setcounter{section}{2}
\section{System Overview}

\begin{figure}[!t]
  \centering
  \includegraphics[width=\linewidth]{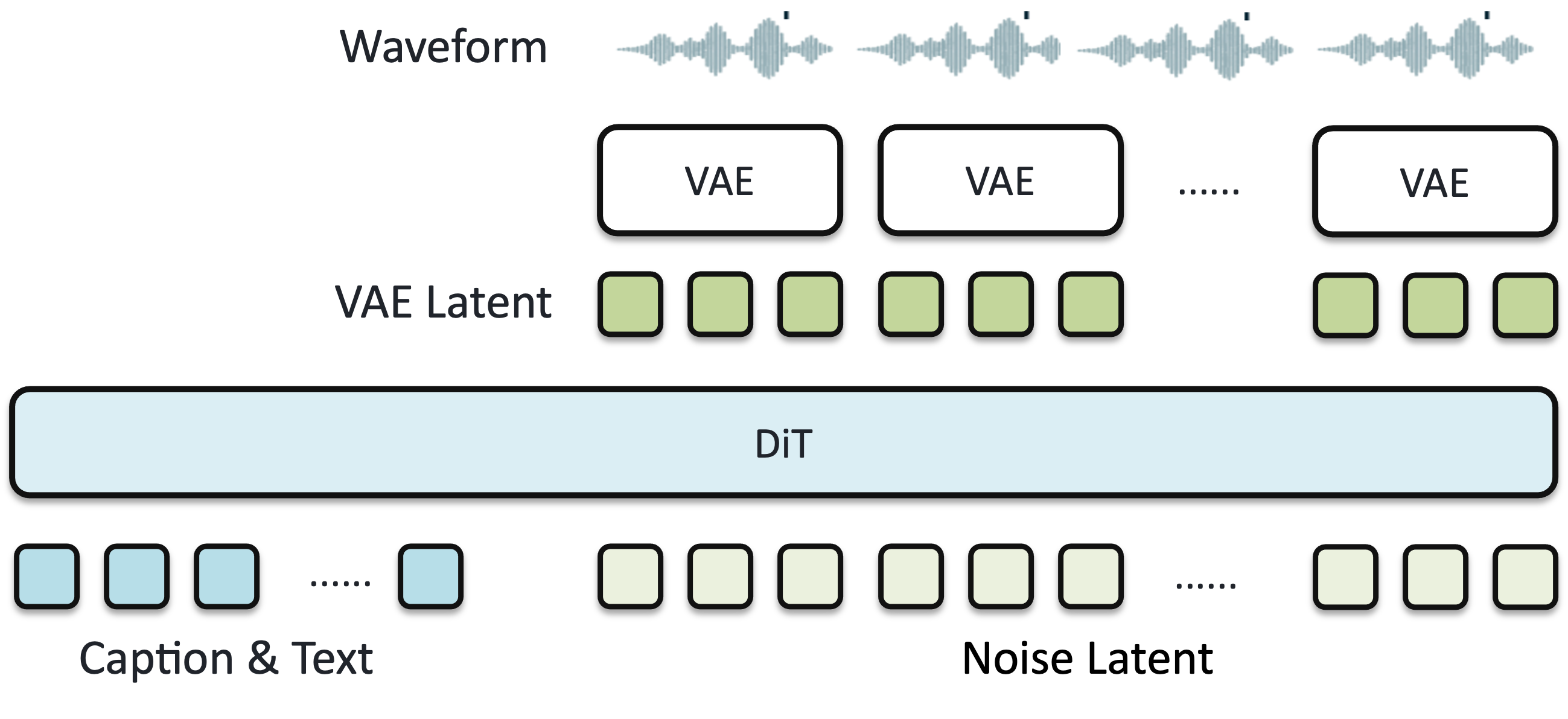}
  \caption{Overview of the unified non-autoregressive audio generation system.
  Caption and text tokens jointly condition a DiT that transforms a
  sequence of noise latents into continuous VAE latents. The shared VAE
  decodes the resulting continuous latent sequence into the complete output
  waveform. The same textual interface, DiT, latent space, and VAE support
  heterogeneous standalone and mixed-scene audio.}
  \label{fig:system-overview}
\end{figure}

The proposed model provides a unified non-autoregressive generation path
for heterogeneous standalone and mixed-scene audio. As illustrated in
Figure~\ref{fig:system-overview}, caption and text tokens jointly condition a
Diffusion Transformer (DiT), which transforms an input sequence of noise
latents into continuous VAE latents for the requested audio. For waveform
reconstruction, the shared VAE decodes the continuous latent sequence generated
by the DiT into the complete output waveform. All supported audio components
share the same conditioning interface, DiT, latent space, and VAE rather than
relying on task-specific generation branches.

The same architecture is used throughout the two-stage training process.
During pre-training, standalone audio samples establish basic acoustic
generation quality and text--audio correspondence. During rich-timeline
supervised fine-tuning, the proposed model learns from multi-character
dialogue, persistent ambience, localized events, long-range structured audio,
and heterogeneous mixtures, while replayed basic samples are used to mitigate
loss of standalone competence. The resulting system uses the same textual
interface, DiT, and VAE for standalone audio, multi-speaker conversations, and
complete scenes containing dialogue, ambience, long-range background structure,
and temporally organized events.

\FloatBarrier

\setcounter{section}{3}
\section{Data}

\subsection{Data Curriculum}
\label{sec:data-curriculum}

A two-stage data curriculum separates broad acoustic learning from supervision
for composing heterogeneous sources in time. Pre-training establishes broad
audio generation quality and text--audio correspondence across standalone
audio. Post-training---the rich-timeline supervised fine-tuning stage
introduced in Section~3---teaches the proposed model how multiple sources are
organized on a shared timeline and rendered within one complete waveform.

The pre-training stage covers linguistic-content audio, long-range structured
audio, and sound-event audio. Linguistic-content audio provides the dominant
supervision mass and supports intelligibility, speaker attributes,
paralinguistic variation, and background acoustics. Long-range structured audio
contributes temporal form, including rhythm, phrasing, instrumentation, and
style continuity. Sound-event audio provides dense supervision for localized
events, acoustic scenes, object interactions, and environmental cues. In
addition to data-family balance, we monitor duration buckets and semantic
coverage during ingestion. Linguistic-content audio is tracked by language and
speaker-related attributes; sound-event audio by event type and acoustic scene;
and long-range structured audio by genre, mood, energy, instrumentation, and
vocal presence.

The post-training stage focuses on organizing multiple sources in a shared
acoustic scene. Its mixture emphasizes long-form dialogue, ambience-rich
mixtures, long-range continuity, and explicit localized events. Long-form
conversational audio provides supervision for speaker consistency and
turn-taking; mixed-scene audio provides foreground/background supervision under
realistic acoustic conditions; long-range structured audio is used to preserve
style continuity and bed-generation ability; and localized sound-event audio
provides supervision for timing and transient-event control.
Table~\ref{tab:data-curriculum} summarizes the relative mixture design. To
mitigate overfitting to noisy or post-produced media, we replay examples of
clean standalone audio during post-training. This replay is intended to
preserve standalone quality while retaining the post-training emphasis on
compositional control.

\begin{table}[H]
  \centering
  \caption{Qualitative composition and role of the data curriculum. Relative
  scales summarize the balance among data families without disclosing exact
  mixture values.}
  \label{tab:data-curriculum}
  \small
  \setlength{\tabcolsep}{4pt}
  \begin{tabular}{>{\raggedright\arraybackslash}p{0.30\linewidth}r>{\raggedright\arraybackslash}p{0.48\linewidth}}
    \toprule
    \textbf{Data family} & \textbf{Relative scale} & \textbf{Main role} \\
    \midrule
    \multicolumn{3}{l}{\textit{Pre-training}} \\
    Linguistic-content audio & Dominant & Linguistic content, speaker attributes, paralinguistics, and background acoustics \\
    Long-range structured audio & Substantial & Long-range form, rhythm, style, instrumentation, and vocal/instrumental attributes \\
    Sound-event audio & Small & Event categories, acoustic scenes, event sources, and environmental context \\
    \midrule
    \multicolumn{3}{l}{\textit{Post-training}} \\
    Long-form conversational audio & Largest & Long-context dialogue, speaker consistency, and turn-taking \\
    Mixed-scene audio & Substantial & Scene realism, ambience, and foreground/background organization \\
    Long-range structured audio & Substantial & Long-range continuity, style control, transitions, and beds \\
    Localized sound-event audio & Small & Foley, transient events, and localized acoustic actions \\
    \bottomrule
  \end{tabular}
\end{table}

Overall, the curriculum progresses from learning what each sound is to learning
how sounds coexist. This section defines the composition of the training data;
the next section describes how heterogeneous examples are converted into
structured annotations.

\FloatBarrier
\subsection{Data Annotation Pipeline}
\label{sec:data-annotation-pipeline}

The data pools described above arrive with heterogeneous audio supervision.
Existing annotations may provide aligned text, source identifiers, language
metadata, captions, catalog-level descriptors, event labels, or acoustic-scene
tags. Long-form media, in contrast, often contain several overlapping
components but only partial source-side annotations. Our annotation pipeline
consolidates the available supervision with audio-derived annotations instead
of reducing every example to a single flat caption. It produces a structured
record that separates persistent scene properties, source-level attributes,
temporally localized content, and primary audible content;
Section~\ref{sec:prompt_enhancement} later describes how this record is
rendered as a textual condition for the proposed model.

\paragraph{Source normalization and temporal decomposition.}
We first normalize source-side annotations into a shared inventory of semantic
fields while preserving the distinction between observed content and descriptive
attributes. Primary audible text is treated as observed content, whereas
language, vocal or acoustic characteristics, environment, style, and production
properties are represented as attributes. Human-readable semantic annotations,
symbolic representations, and learned acoustic representations are stored
separately. Learned representations, including discrete audio tokens and
continuous VAE latents, retain acoustic detail but are not assumed to be
serialized into natural-language templates. For long-form audio, annotation is
performed on a shared time axis. Source activity, persistent soundscape
segments, dialogue turns, and candidate localized events are identified
independently, so their intervals may overlap. This is important for mixed
media, where a dialogue turn can coexist with persistent ambience, background
source activity, and a short foreground event.

\paragraph{Dialogue and role linking.}
Dialogue activity is segmented into utterances, and available primary audible
text is aligned to the corresponding intervals. Speaker diarization then
assigns utterances to local speaker tracks. Tracks attributed to the same source
are linked across dialogue turns to form a stable role identifier. Vocal
attributes are aggregated at the role level rather than predicted independently
for each utterance. A role profile can include identity information, gender or
voice class, age when reliably available, accent or dialect, and timbre
descriptors. This aggregation reduces turn-level inconsistency and provides the
anchors needed to associate every dialogue line with the correct voice over a
long context.

\paragraph{Soundscape annotation.}
Persistent acoustic conditions are annotated separately from local events. At
the clip or scene level, the annotation describes the environment, room tone,
background activity, and spatial or production characteristics that remain
meaningful across multiple intervals. For mixed media, source entry, exit, and
transition regions may additionally be localized when they form part of the
temporal organization of the scene. These entries describe how persistent
source activity is placed within a mixed soundscape.

\paragraph{Localized sound events.}
Transient and independently meaningful sounds are retained as time-localized
events even when they overlap with dialogue, ambience, or other persistent
source activity. Each event is associated with an interval or temporal anchor
and, when supported by the audio, a description of the event, its acoustic
source, and the underlying action. Event annotations are not absorbed into a
dialogue caption or a global scene description. Repeated detections of the same
acoustic occurrence are merged, while distinct overlapping events remain
separate entries on the shared timeline.

\paragraph{Hierarchical assembly and consistency checks.}
The component annotations are finally reconciled into a structured record,
\begin{equation}
    R = (G, \mathcal{P}, \mathcal{E}, U),
    \label{eq:structured-record}
\end{equation}
where $G$ is the global scene and soundscape, $\mathcal{P}$ is an optional set
of source or character profiles, $\mathcal{E}=\{e_i\}_{i=1}^N$ is a
time-ordered sequence of local events, and $U$ is the primary audible content.
This assembly stage resolves duplicated or conflicting descriptions, binds
dialogue turns to role profiles, and determines whether a sound is persistent
context or a localized occurrence. We apply structural consistency checks before admitting
an example to the rich-timeline pool. Every visible role identifier must resolve
to one profile; dialogue, primary audible content, and localized events must
remain aligned to their audible intervals; event ordering must agree with the
waveform; and overlapping components must not be removed merely because another
source is dominant.
Attributes that cannot be assigned to a reliable owner or temporal scope are
omitted rather than attached speculatively.

Real recordings therefore produce the structured record $R$. The synthetic
recipes in Section~\ref{sec:synthetic-audio-data} are mapped to compatible
records, and Section~\ref{sec:prompt_enhancement} organizes these records into
the textual conditions used by the proposed model.
\FloatBarrier

\subsection{Synthetic Audio Data Construction}
\label{sec:synthetic-audio-data}

Although the annotation pipeline above recovers structured supervision from
real recordings, such data rarely provide independent control over event count,
onset, duration, overlap, relative level, and spatial attributes. We therefore
construct a complementary synthetic corpus using a recipe-driven soundscape
synthesis framework. The design builds on the
controllable foreground/background composition paradigm introduced by
Scaper~\cite{salamon2017scaper}, subsequently adopted by DESED to create
strongly annotated synthetic soundscapes~\cite{turpault2019desed} and used by
FUSS to construct source-preserving mixtures with simulated room
responses~\cite{wisdom2021fuss}. Rather than rendering a scene as an
indivisible waveform, we represent it as a collection of independently
addressable sources. The recipe exposes explicit parameters for acoustic
content, temporal organization, spatial rendering, and signal degradation.

\paragraph{Source inventory.}
The source pool comprises screened heterogeneous vocal, foreground,
background, and ambient audio sources. These signals are treated as authoring
components: \emph{foreground} and \emph{background} describe a source's role
within a particular recipe rather than an intrinsic property of the recording.
Parameterized procedural synthesis supplies additional primitives, including
sinusoids, frequency sweeps, clicks, metronomes, beeps, impulses, and
band-limited noise. For rare or designed sounds, such as fictional machines or
magical events, we layer physically interpretable recordings with procedural
components and, when necessary, quality-screened candidates produced by audio
generation models.

\paragraph{Scene specification and temporal scheduling.}
Each example is instantiated from a scene specification that assigns every
source a semantic role, source identifier, onset, duration, and relative level.
At the scene level, the specification defines event counts, ordering and overlap
constraints, and target-to-background signal-to-noise ratios. Given a recipe
and random seed, a deterministic scheduler realizes exact repetition patterns,
silence intervals, tempo constraints, causal event chains, and long-form state
transitions. The same representation supports
object--material--action contrasts, dense polyphonic scenes, and multi-stage
narratives without collapsing their constituent events into a single
uncontrolled generation step.

\paragraph{Acoustic and channel rendering.}
Acoustic variation is applied independently to each source. Depending on the
recipe, a source may be rendered with room or binaural impulse responses and
conditioned on direction, distance, motion trajectory, occlusion, and
propagation effects. Our parameterized spatial simulation follows the general
design principles of SpatialScaper~\cite{roman2024spatialscaper}. Device and
channel conditions are implemented as ordered processing chains that may
include band limitation, equalization, static or background noise, nonlinear
distortion, resampling, codec processing, and packet-loss effects. Whenever
available, we retain both the clean input and each processed source track in
addition to the complete mixed waveform.

\paragraph{Controlled counterfactuals.}
We construct paired counterfactual examples by cloning a valid scene recipe,
modifying exactly one factor, and freezing all remaining factors. An
intervention may change the number of events, exchange their order, remove a
target, insert an additional event, or alter material, direction, distance, or
channel conditions. Each intervention updates the recipe and its condition
before the corresponding target waveform is rendered, producing a new valid
condition--target pair. The resulting pairs reduce nuisance variation and
support controlled positive, negative, and hard-negative comparisons for
individual control dimensions. This construction differs from the conditional
dropout in Section~\ref{sec:semantic-conditional-views}, which hides condition
content without changing the current sample's true event order or target audio.

\paragraph{Annotations, quality control, and limitations.}
For every rendered mixture, we retain the instantiated recipe, random seed,
component tracks or processed source signals, and construction-time event
annotations. The recipe and construction-time event annotations are mapped into
the structured record defined in Section~\ref{sec:data-annotation-pipeline}. We
refer to their temporal
boundaries as \emph{construction-time strong labels}, since a
boundary specified by a recipe need not coincide perfectly with the
perceptually audible activity in the corresponding source recording. Quality
control comprises checks of semantic content, aligned textual content when
applicable, event presence and count, temporal order, spectrum, loudness, true
peak, clipping, aliasing, discontinuities, and rendering artifacts, followed by
listening review of sampled outputs. Loudness and true-peak measurements follow ITU-R
BS.1770-5~\cite{itu2023bs1770}. Synthetic scenes complement rather than replace
real recordings: residual context in source clips and mismatch between
simulated and real acoustics remain important limitations, and any downstream
benefit of synthetic data must be established empirically.
Together, real annotations and synthetic recipes provide compatible structured
records for training-side conditioning. Section~\ref{sec:prompt_enhancement}
also converts inference-side user input into this shared format before rendering
the textual condition for the proposed model.

\subsection{Prompt Enhancement and Condition Rendering}
\label{sec:prompt_enhancement}

Real-world user prompts are often free-form and may describe standalone audio,
character dialogue, ambience, foreground events, long-range structured content,
or mixed scenes. Each request can also contain detailed requirements, such as
source identity, acoustic attributes, performance cues, spatial setting, and
temporal relationships. Directly using such prompts may cause the proposed
model to miss important details or misunderstand the relationships among audio
components. We therefore develop a prompt enhancement (PE) module that maps
free-form user inputs to the structured fields required by the condition
renderer while preserving the user's explicit requirements and the stated
relationships among sound events.

We adopt a large language model to organize each user request into a unified
set of condition fields. The PE module extracts the global scene, ambience, and
production context; identifies source or role profiles together with voice or
acoustic attributes and performance cues; preserves user-provided primary
audible text verbatim, such as a transcript or lyrics when explicitly supplied;
extracts localized sound events, spatial attributes, and sound evolution; and
estimates start times, end times, ordering, interruptions, and overlap
relations. It does not introduce roles, dialogue lines, sources, or events that
the user did not request.

\paragraph{Condition rendering.}
The annotation pipeline, synthetic construction process, and PE module all
produce records compatible with $R$ in Equation~\ref{eq:structured-record}. To
form the input to the proposed model, we first select a surface template from a finite family and
apply the renderer $\mathcal{T}$,
\begin{equation}
    C_{\mathrm{full}} = \mathcal{T}(R).
    \label{eq:full-condition}
\end{equation}
Given a structured record and the selected template, the renderer is
deterministic. Its stable prompt grammar places the global scene and ambience
before source or role profiles, followed by time-sorted primary audible content
and localized events. Primary audible text remains explicit and distinct from
style and production attributes. $C_{\mathrm{full}}$ denotes the complete
textual condition and may contain both a scene caption and primary audible
content fields.

The renderer also enforces a token budget, omitting low-priority detail by rule
rather than truncating the tail of an arbitrary prompt. Variation within the
finite template family changes surface wording without hallucinating conditions
or changing the underlying record.

\paragraph{Validation and repair.}
The PE output is validated against the template rules required by the proposed
model. The validation checks whether the user's requirements are preserved, the
output format is complete, roles and sources are correctly bound to their
audible content, user-provided audible text remains verbatim, localized events
follow the intended order, and timeline and duration constraints are satisfied.
If the rewritten result fails these checks, the module performs a targeted
repair using the reported validation errors. After validation, the renderer
returns the enhanced textual condition, including the estimated timeline, as
input to the proposed model.

Training annotations, synthetic records, and PE outputs therefore use
compatible condition formats. Section~5 describes how different conditional
views are constructed from this format during training.

\section{Conditional Training for Rich-Timeline Audio}
\subsection{Semantic Conditional Views}
\label{sec:semantic-conditional-views}

Rich-timeline training uses alternative condition views derived from the
structured record in Equation~\ref{eq:structured-record} and the renderer in
Section~\ref{sec:prompt_enhancement}. This section specifies which semantic
units remain visible in each view.

In rich-timeline audio, independently dropping individual fields can leave
dialogue as the dominant explanation for the target audio and weaken the
contrast supplied by scene conditions. We therefore apply dropout to structured
semantic units before rendering:
\begin{equation}
    C_M = \mathcal{T}(\mathcal{D}_M(R)),
    \qquad
    M\in\{\mathrm{full},\mathrm{dialogue},\mathrm{scene},\varnothing\}.
    \label{eq:semantic-views}
\end{equation}
Here, $\mathcal{D}_M$ hides semantic units in $R$ before $\mathcal{T}$ renders
the condition. The full view retains every available condition. The dialogue
view retains dialogue, role identity, and the necessary attributes needed to
interpret them. The scene view retains the persistent soundscape and
independent sound events while removing dialogue and roles together. The empty
view is the unconditional condition. The mixture over these views is
configurable, but each selected view must remain an interpretable request for
the target audio.

Fine-grained dropout follows the same rule. An independent event may lose its
description or timestamp only if the remaining row still provides an
interpretable event description or localization; otherwise the complete event
row is removed. After any
deletion, relations cannot refer to a removed event or source. Fields whose
semantics cannot be separated safely remain in the full view. Conditional
dropout hides only condition content: it does not rewrite primary audible text,
permute the true event order, or change the target waveform.

\subsection{Attribute Contrast and Role-Bundle Integrity}
\label{sec:attribute-role-integrity}

Attribute contrast and long-context role binding place different constraints
on conditional dropout. We apply each rule only where the remaining condition
is semantically interpretable.

\paragraph{Attribute contrast.}
Age, gender or voice class, accent, and dialect use a canonical attribute block
with an elevated and configurable missing probability relative to decorative
detail. Across repeated draws, an independently interpretable attribute can be
present or absent for the same target audio. This contrast is intended to expose
the attribute-specific conditional difference to CFG. Whenever an attribute is
visible, it remains normalized, explicit, and unambiguous; the sampling
distribution is configured to retain attribute-present examples as well as
attribute-absent draws.

\paragraph{Role-bundle integrity.}
A rich-timeline role bundle that is referenced by visible dialogue cannot be
partially deleted. Whenever a visible turn is labeled \texttt{roleX}, its role
card retains every available core anchor: identity, gender or voice class, age
when annotated, accent or dialect when annotated, and at least one timbre
anchor. Optional personality or lengthy descriptive prose may be dropped. The
role card and its dialogue turns remain atomically bound, so deleting a role
bundle also removes its linked turns from the visible condition. When reference
audio is present, the reference audio, its textual label, and the matching label
in the target dialogue remain bound.

Thus, independently interpretable attributes can participate in attribute
contrast, whereas a role bundle already referenced by dialogue can only be
removed together with its linked turns.

Counterfactual construction in Section~\ref{sec:synthetic-audio-data} creates a
new valid condition--target pair; conditional dropout instead leaves the
audible content of the fixed pair unchanged.

\subsection{Classifier-Free Guidance}
\label{sec:classifier-free-guidance}

The semantic views in Section~\ref{sec:semantic-conditional-views} provide the
visible and empty conditions used for classifier-free guidance (CFG). At
inference, CFG combines the corresponding conditional and unconditional vector
fields as
\begin{equation}
    v_{\mathrm{cfg}} = v_u + s\bigl(v_c-v_u\bigr),
    \label{eq:cfg}
\end{equation}
where $v_c$ is predicted from the visible condition, $v_u$ is predicted from
the empty condition, and $s$ is the guidance scale. The difference $v_c-v_u$
represents the proposed model's learned response to information in the visible
condition, and $s$ weights that difference. In this formulation, guidance relies
on contrasts learned from consistent views and does not itself supply role or
soundscape information absent from the visible condition.

\subsection{Shared Continuous Audio VAE}
\label{sec:audio-vae}

The conditional mechanisms described above operate on a continuous audio
latent space, which is defined for all supported domains by the shared VAE
described here.

\paragraph{Design motivation.}
Qwen-Audio-3.0-Gen-Preview covers heterogeneous standalone and mixed-scene
audio and therefore requires a shared latent representation that supports
high-fidelity reconstruction across these audio settings. We compress 48\,kHz stereo waveforms
into a continuous latent sequence at 25\,Hz, reducing the sequence length
processed by the DiT. We first train the VAE for high-quality reconstruction
and subsequently introduce auxiliary semantic supervision so that the latent
representation contains semantic information and is intended to better support
downstream audio generation.

\paragraph{Latent representation.}
Following the convolutional waveform autoencoder design of Stable Audio
Open~\cite{evans2025sao}, our VAE maps each waveform to a 128-dimensional
continuous latent sequence. The encoder parameterizes a diagonal Gaussian
posterior, and the decoder reconstructs the waveform from a sampled latent:
\begin{equation}
\begin{aligned}
q_{\phi}(\mathbf{z}\mid\mathbf{x})
    &=
    \mathcal{N}\!\left(
        \boldsymbol{\mu}_{\phi}(\mathbf{x}),
        \operatorname{diag}
        \boldsymbol{\sigma}_{\phi}^{2}(\mathbf{x})
    \right), \\
\widehat{\mathbf{x}}
    &=
    D_{\theta}(\mathbf{z}),
    \qquad
    \mathbf{z}\sim q_{\phi}(\mathbf{z}\mid\mathbf{x}).
\end{aligned}
\label{eq:vae-overview}
\end{equation}
The latent sample $\mathbf{z}$ is used for waveform reconstruction, whereas the
posterior mean $\boldsymbol{\mu}_{\phi}(\mathbf{x})$ is used for semantic
supervision.

\paragraph{Semantic supervision.}
The posterior mean is mapped by a lightweight projection module into the
embedding space of a frozen Qwen2.5-3B base language model. The projected audio
representations are combined with an instruction prompt, and the frozen
language model supplies a next-token prediction objective over the target
response. Gradients update the VAE encoder and projection module, while the
language-model parameters remain fixed. The language model is used only as an
auxiliary supervisor during VAE training and is absent from downstream audio
generation.

\paragraph{Training objective.}
The VAE generator objective combines reconstruction, variational, adversarial,
feature-matching, and semantic terms:
\begin{equation}
\mathcal{L}_{G}
=
\mathcal{L}_{\mathrm{rec}}
+
\lambda_{\mathrm{KL}}\mathcal{L}_{\mathrm{KL}}
+
\mathbb{I}_{\mathrm{adv}}
\left(
    \lambda_{\mathrm{adv}}\mathcal{L}_{\mathrm{adv}}
    +
    \lambda_{\mathrm{fm}}\mathcal{L}_{\mathrm{fm}}
\right)
+
\mathbb{I}_{\mathrm{sem}}
\lambda_{\mathrm{sem}}\mathcal{L}_{\mathrm{sem}},
\label{eq:vae-objective}
\end{equation}
where $\mathcal{L}_{\mathrm{rec}}$ contains multi-resolution spectral
reconstruction terms evaluated on stereo waveform representations,
$\mathcal{L}_{\mathrm{KL}}$ regularizes the variational posterior, and
$\mathcal{L}_{\mathrm{adv}}$ and $\mathcal{L}_{\mathrm{fm}}$ denote the
adversarial and discriminator feature-matching objectives. The indicators
$\mathbb{I}_{\mathrm{adv}}$ and $\mathbb{I}_{\mathrm{sem}}$ specify whether the
corresponding objectives are active for the current training stage and sample.

\paragraph{VAE training stages.}
We adopt a progressive training strategy distinct from the proposed model's
pre-training and post-training curriculum in
Section~\ref{sec:data-curriculum}. The VAE is trained on a large-scale in-house
audio corpus. We first optimize the waveform VAE with
reconstruction, KL, adversarial, and feature-matching objectives to obtain a
high-fidelity acoustic representation. Starting from this checkpoint, semantic continuation
adds the frozen language model and projection module. The discriminator is
temporarily disabled during early continuation; reconstruction-only and
semantically annotated samples are jointly used, with the semantic objective
applied only to the latter. The discriminator is then reintroduced while the
reconstruction and semantic objectives are jointly optimized. This schedule is
designed to incorporate semantic information while limiting degradation of the
acoustic representation learned during reconstruction training.

Section~\ref{sec:evaluation} first evaluates the complete system on audio
benchmarks and then isolates the VAE through reconstruction measurements and a
controlled downstream probe.

\section{Evaluation}
\label{sec:evaluation}

We evaluate the proposed model across two broad evaluation settings:
reference-conditioned generation and text-conditioned audio generation. The
former covers zero-shot generation and multi-speaker consistency; the latter
covers caption fulfillment, perceptual quality, rich-timeline control, and
long-range structure. We use public benchmarks when available and in-house
benchmarks for multi-speaker and temporal capabilities not directly covered by
them. Seed-TTS-Eval~\cite{a2024seedtts}
serves as the public basis for the reference-conditioned tests, and the in-house
multi-speaker cases are drawn from its English and Chinese test data. In each
zero-shot instance, the proposed model receives a reference utterance and target
transcript and must synthesize the target content while preserving the reference
voice.


\begin{table}[H]
\caption{
Results on the Seed-TTS-Eval benchmark.
WER/CER (\%) measure content intelligibility, and SIM denotes WavLM
speaker similarity on a $[0,1]$ scale.
Bold and underlined values indicate the best and second-best results among the
evaluated models in each column, respectively.
A dash indicates that a result is unavailable or not applicable.
Beyond TTS marks support for audio generation, interaction, understanding, or
editing beyond conventional speech synthesis.
}
\label{tab:seed_tts_eval}
\centering
\setlength{\tabcolsep}{3.5pt}
\resizebox{\linewidth}{!}{%
\begin{tabular}{lccccccc}
\toprule
\multirow{2}{*}{\textbf{Model}}
& \multicolumn{2}{c}{\textbf{EN}}
& \multicolumn{2}{c}{\textbf{ZH}}
& \multicolumn{2}{c}{\textbf{Hard-ZH}}
& \multirow{2}{*}{\shortstack{\textbf{Beyond}\\\textbf{TTS}}} \\
\cmidrule(lr){2-3}
\cmidrule(lr){4-5}
\cmidrule(lr){6-7}
& \textbf{WER (\%)} $\downarrow$
& \textbf{SIM} $\uparrow$
& \textbf{CER (\%)} $\downarrow$
& \textbf{SIM} $\uparrow$
& \textbf{CER (\%)} $\downarrow$
& \textbf{SIM} $\uparrow$
& \\
\midrule

Human
& 2.14
& 0.734
& 1.26
& 0.755
& --
& --
& -- \\

\midrule
\multicolumn{8}{c}{\textbf{Autoregressive Models}} \\
\midrule

Seed-TTS \cite{a2024seedtts}
& 2.25
& 0.762
& 1.12
& 0.796
& 7.59
& 0.776
& -- \\

Ming-UniAudio \cite{yan2025minguniaudio}
& 1.85
& 0.580
& 0.95
& 0.700
& --
& --
& $\checkmark$ \\

MiMo-Audio-7B-Instruct \cite{coreteam2025mimo}
& 5.37
& --
& 1.96
& --
& 14.14
& --
& $\checkmark$ \\

UniMoE-Audio \cite{liu2025unimoeaudio}
& 1.90
& 0.560
& \underline{0.80}
& 0.650
& --
& --
& $\checkmark$ \\

Step-Audio-2-Mini \cite{wu2025stepaudio2}
& 3.18
& --
& 2.13
& --
& 16.31
& --
& $\checkmark$ \\

Qwen3.5-Omni-Plus \cite{qwenteam2026qwen35omni}
& \underline{1.26}
& --
& 0.99
& --
& --
& --
& $\checkmark$ \\

VibeVoice-1.5B \cite{peng2026vibevoice}
& 3.04
& 0.689
& 1.16
& 0.744
& --
& --
& -- \\

FireRedTTS-2 \cite{xie2025fireredtts2}
& 1.95
& 0.665
& 1.14
& 0.736
& --
& --
& -- \\

Qwen3-TTS-12Hz-1.7B-Base
\cite{hu2026qwen3tts}
& \textbf{1.24}
& --
& \textbf{0.77}
& --
& --
& --
& -- \\

MiniMax-Speech
\cite{zhang2025minimaxspeech}
& 1.90
& 0.738
& 0.99
& 0.799
& --
& --
& -- \\

VoxCPM2 \cite{zhou2026voxcpm2}
& 1.84
& 0.753
& 0.97
& 0.795
& 8.13
& 0.753
& -- \\

CosyVoice3-1.5B
\cite{du2025cosyvoice3}
& 2.21
& 0.720
& 1.12
& 0.781
& \textbf{5.83}
& 0.758
& -- \\

Qwen-Audio-3.0-TTS \cite{qwenaudio30ttsfreelycontrollablehighly}
& 1.54
& 0.762
& 0.84
& 0.792
& 7.00
& 0.768
& -- \\

\midrule
\multicolumn{8}{c}{\textbf{Non-Autoregressive Models}} \\
\midrule

F5-TTS (32 NFE)
\cite{chen2025f5}
& 1.83
& 0.647
& 1.56
& 0.741
& 8.67
& 0.713
& -- \\

F5R-TTS \cite{sun2025f5rtt}
& --
& --
& 1.37
& 0.754
& 8.79
& 0.718
& -- \\

LongCat-AudioDiT-3.5B
\cite{xin2026longcataudiodit}
& 1.50
& \underline{0.786}
& 1.09
& \underline{0.818}
& \underline{6.04}
& \underline{0.797}
& -- \\

\midrule

\textbf{Qwen-Audio-3.0-Gen-Preview}
& 1.61
& \textbf{0.805}
& 1.06
& \textbf{0.819}
& 8.25
& \textbf{0.808}
& $\checkmark$ \\

\bottomrule
\end{tabular}%
}
\end{table}

Qwen-Audio-3.0-Gen-Preview records the highest SIM on the EN, ZH, and
Hard-ZH subsets, making speaker preservation its clearest numerical strength
in this evaluation. Its WER/CER values are not the column minima, so the results
do not establish a uniform advantage in linguistic accuracy. The proposed
model also supports the broader Beyond TTS task range, showing that a
unified generation scope is compatible with strong zero-shot speaker
similarity without implying leadership on every TTS metric.


The public benchmark evaluates one target speaker per instance. To assess joint
control over multiple speakers, we construct an in-house benchmark for
reference-conditioned multi-speaker audio generation.

The benchmark contains a few hundred instances across English and Chinese, with
roughly a hundred distinct reference speakers in each language. Each instance contains $K$ reference
utterances from distinct speakers and a speaker-attributed target
script $\{(s_i, t_i)\}_{i=1}^{N}$, where $s_i$ indicates the intended
reference speaker and $t_i$ is the text of the $i$-th turn. The proposed model
is required to render the entire dialogue as a single waveform while
assigning each turn to the correct reference voice and preserving
speaker identity across turns.

\begin{table}[H]
\caption{Results on our reference-conditioned multi-speaker benchmark. WER and CER (\%) measure content intelligibility over the complete dialogue waveform. SIM denotes WavLM speaker similarity between each generated segment and its reference audio. Consistency (CONS) denotes WavLM speaker similarity between segments of the same speaker across dialogue turns, reflecting voice preservation in multi-turn generation. Both similarity metrics use a $[0,1]$ scale. Bold and underlined values indicate the best and second-best results in each column, respectively.}
\label{tab:multispeaker_tts}
\centering
\setlength{\tabcolsep}{3pt}
\resizebox{\linewidth}{!}{%
\begin{tabular}{lcccccc}
\toprule
\multirow{2}{*}{\textbf{Model}}
& \multicolumn{3}{c}{\textbf{EN}}
& \multicolumn{3}{c}{\textbf{ZH}} \\
\cmidrule(lr){2-4}
\cmidrule(lr){5-7}
& \textbf{WER (\%)} $\downarrow$
& \textbf{SIM} $\uparrow$
& \textbf{CONS} $\uparrow$
& \textbf{CER (\%)} $\downarrow$
& \textbf{SIM} $\uparrow$
& \textbf{CONS} $\uparrow$ \\
\midrule
Seed-Audio-1.0
& \textbf{1.20}
& \textbf{0.575}
& \underline{0.659}
& \textbf{1.35}
& \underline{0.661}
& \underline{0.704} \\
\textbf{Qwen-Audio-3.0-Gen-Preview}
& \underline{1.32}
& \underline{0.514}
& \textbf{0.702}
& \underline{1.99}
& \textbf{0.682}
& \textbf{0.740} \\
\bottomrule
\end{tabular}%
}
\end{table}

Compared with Seed-Audio-1.0\footnote{\url{https://docs.byteplus.com/en/docs/byteplusvoice/seedaudio-01}},
Qwen-Audio-3.0-Gen-Preview records higher CONS in both languages and
higher SIM in ZH. The baseline records lower EN WER and ZH CER, as well as
higher EN SIM. Thus, the proposed model's consistent advantage across both
languages lies in cross-turn speaker consistency, rather than a uniform lead in
intelligibility or speaker fidelity.

We next turn from reference-conditioned generation to broader text-conditioned
audio generation. On AudioCaps, we evaluate audio generation on the test set
\cite{kim-etal-2019-audiocaps}, using the publicly accessible OpenSound
release, which contains 4,411 test caption instances.\footnote{
\url{https://huggingface.co/datasets/OpenSound/AudioCaps/viewer/default/test}}
Because multiple audio realizations may satisfy a caption, we use complementary
automatic alignment measures rather than direct waveform-reference metrics.
All systems are evaluated using the same set of caption prompts. We report
LAION-CLAP and MS-CLAP on the full test set to maintain comparability with
established AudioCaps evaluations. We additionally employ
Qwen3-Omni-30B-A3B-Instruct~\cite{xu2025qwen3omnitechnicalreport} as an audio-language judge over the full set.

To provide a complementary cross-model evaluation, we uniformly sample
a fixed subset on the order of a hundred prompts without replacement using a
fixed random seed.
The same prompts and generated audio samples are evaluated using both CLAP metrics
and three independent large audio-language model (LALM) judges:
Qwen3-Omni-30B-A3B-Instruct, Qwen3.5-Omni-Plus\footnote{\url{https://help.aliyun.com/en/model-studio/qwen-omni}}, and
Gemini-3.1-Pro-Preview\footnote{\url{https://ai.google.dev/gemini-api/docs/models/gemini-3.1-pro-preview}}. The judges assess whether the generated audio
satisfies the events, attributes, and temporal relations specified by the
input caption. All baseline systems are reproduced following the settings
described in their original papers, and the reproduced CLAP results are
consistent with their reported performance.

\begin{table*}[t]
\caption{
Results on the AudioCaps benchmark. The full-set evaluation contains
4,411 caption prompts, while the fixed random subset is on the order of a
hundred prompts sampled uniformly without replacement using a fixed seed.
Qwen3 denotes Qwen3-Omni-30B-A3B-Instruct, Qwen3.5 denotes
Qwen3.5-Omni-Plus, and Gemini denotes Gemini-3.1-Pro-Preview.
All LALM columns report mean audio--text semantic-alignment scores.
Bold and underlined values indicate the best and second-best results in
each column, respectively.
}
\label{tab:audiocaps}
\centering
\small
\setlength{\tabcolsep}{3.6pt}
\resizebox{\textwidth}{!}{%
\begin{tabular}{lcccccccc}
\toprule
\multirow{3}{*}{\textbf{Model}}
& \multicolumn{3}{c}{\textbf{Full Test Set} ($n=4{,}411$)}
& \multicolumn{5}{c}{\textbf{Fixed Random Subset}} \\
\cmidrule(lr){2-4}
\cmidrule(lr){5-9}
& \multicolumn{2}{c}{\textbf{CLAP} $\uparrow$}
& \multicolumn{1}{c}{\textbf{LALM} $\uparrow$}
& \multicolumn{2}{c}{\textbf{CLAP} $\uparrow$}
& \multicolumn{3}{c}{\textbf{LALM} $\uparrow$} \\
\cmidrule(lr){2-3}
\cmidrule(lr){4-4}
\cmidrule(lr){5-6}
\cmidrule(lr){7-9}
& \textbf{LAION}
& \textbf{MS}
& \textbf{Qwen3}
& \textbf{LAION}
& \textbf{MS}
& \textbf{Qwen3}
& \textbf{Qwen3.5}
& \textbf{Gemini} \\
\midrule
AudioX \cite{tian2025audiox}
& 0.3208
& 0.4042
& \underline{0.7832}
& 0.3137
& 0.3980
& \underline{0.7995}
& 0.7000
& 0.8400 \\

UniFlow-Audio \cite{xu2025uniflow}
& \textbf{0.3712}
& 0.3924
& 0.7467
& \textbf{0.3765}
& 0.3852
& 0.7245
& \underline{0.8100}
& \textbf{0.9185} \\

Dasheng-AudioGen \cite{mei2026dashengaudiogen}
& 0.1969
& \textbf{0.5086}
& 0.7327
& 0.2003
& \textbf{0.5053}
& 0.7325
& 0.6790
& 0.7625 \\

MMAudio \cite{cheng2025mmaudio}
& \underline{0.3612}
& \underline{0.4317}
& 0.7095
& \underline{0.3618}
& \underline{0.4416}
& 0.6775
& 0.7215
& 0.8375 \\
\midrule
\textbf{Qwen-Audio-3.0-Gen-Preview}
& 0.3178
& 0.3732
& \textbf{0.8393}
& 0.3002
& 0.3549
& \textbf{0.8375}
& \textbf{0.8205}
& \underline{0.8540} \\
\bottomrule
\end{tabular}%
}
\end{table*}

\begin{table}[t]
\caption{
Perceptual-quality results on the AudioCaps benchmark evaluated using the
AudioBox metrics PQ, PC, CE, and CU. All systems are evaluated on the same test
prompts.
Bold and underlined values indicate the best and second-best results in
each column, respectively.
}
\label{tab:audiocaps_audiobox}
\centering
\small
\setlength{\tabcolsep}{6pt}
\begin{tabular}{lcccc}
\toprule
\textbf{Model}
& \textbf{PQ} $\uparrow$
& \textbf{PC} $\uparrow$
& \textbf{CE} $\uparrow$
& \textbf{CU} $\uparrow$ \\
\midrule
AudioX \cite{tian2025audiox}
& 5.711
& \underline{3.196}
& \underline{3.528}
& 4.969 \\

UniFlow-Audio \cite{xu2025uniflow}
& 5.647
& 3.125
& 3.435
& 4.923 \\

Dasheng-AudioGen \cite{mei2026dashengaudiogen}
& \underline{5.942}
& 2.948
& 3.493
& \underline{5.180} \\

MMAudio \cite{cheng2025mmaudio}
& 5.493
& 2.989
& 3.309
& 4.867 \\
\midrule
\textbf{Qwen-Audio-3.0-Gen-Preview}
& \textbf{6.256}
& \textbf{3.616}
& \textbf{3.960}
& \textbf{5.389} \\
\bottomrule
\end{tabular}
\end{table}

In Table~\ref{tab:audiocaps}, Qwen-Audio-3.0-Gen-Preview leads the
full-set Qwen3 column and the subset Qwen3 and Qwen3.5 columns, ranks second
under Gemini, and does not lead any CLAP column. Its mean across the three
subset judges is 0.8373, compared with the next-highest mean of 0.8177 from
UniFlow-Audio; its full-set and subset Qwen3 scores differ by 0.0018.

The CLAP and LALM results exhibit different system-level rankings,
reflecting their distinct evaluation mechanisms. At the individual-example
level, the Spearman correlations between the three LALM judges and
LAION-CLAP range from 0.136 to 0.254, while those with MS-CLAP range from
0.037 to 0.091. These low correlations support treating the LALM judgments as
complementary to, rather than replacements for, the established embedding-based
CLAP measures: the former directly assess fulfillment of events, attributes,
and temporal relations in the caption.

Table~\ref{tab:audiocaps_audiobox} further evaluates perceptual quality
using AudioBox~\cite{tjandra2025audiobox}.
Qwen-Audio-3.0-Gen-Preview records the highest values across all four
dimensions. Taken together, the results locate its clearest numerical
advantages in the LALM and AudioBox assessments rather than CLAP, and do not
establish a uniform advantage across evaluation mechanisms.

On the in-house rich-timeline audio benchmark, we evaluate compositional and
temporal instruction following. Global embedding-based metrics on benchmarks such as AudioCaps provide an
aggregate measure of audio--text alignment, but do not explicitly assess
whether the requested events occur at the specified times. We therefore
construct an in-house benchmark at a scale on the order of a hundred
rich-timeline audio scenes, with durations on the order of tens of seconds, to evaluate event coverage and
temporal adherence. Each prompt specifies multiple audio events
together with their start and end times.

For each generated sample, Gemini-3.1-Pro-Preview and
Qwen3.5-Omni-Plus independently determine whether each requested event is
audible and estimate its temporal interval. We report the proportion of
requested events judged audible (event recall), mean intersection over union
(mIoU) between each reference event and its
best-matched predicted interval, and the proportions of events reaching IoU
thresholds of 0.3 and 0.5 in Table~\ref{tab:t2a_temporal}.

\begin{table}[H]
\caption{Results on our in-house rich-timeline audio benchmark for
compositional and temporal instruction following under Gemini-3.1-Pro-Preview
and Qwen3.5-Omni-Plus judges. All values are percentages. Bold and underlined
values indicate the best and second-best results under each judge,
respectively.}
\label{tab:t2a_temporal}
\centering
\small
\setlength{\tabcolsep}{6pt}
\renewcommand{\arraystretch}{1.12}
\begin{tabular*}{\textwidth}{
    @{\extracolsep{\fill}}
    l
    cccc
    @{}
}
\toprule
\textbf{Model}
& \textbf{Recall} $\uparrow$
& \textbf{mIoU} $\uparrow$
& \textbf{IoU@.3} $\uparrow$
& \textbf{IoU@.5} $\uparrow$ \\
\midrule

\multicolumn{5}{@{}l}{\textit{Gemini-3.1-Pro-Preview as Judge}} \\
\addlinespace[1pt]

Seed-Audio-1.0
& \textbf{98.77}
& \underline{38.48}
& \underline{56.17}
& \underline{38.27} \\
\textbf{Qwen-Audio-3.0-Gen-Preview}
& \underline{88.58}
& \textbf{43.73}
& \textbf{66.36}
& \textbf{50.00} \\

\midrule

\multicolumn{5}{@{}l}{\textit{Qwen3.5-Omni-Plus as Judge}} \\
\addlinespace[1pt]

Seed-Audio-1.0
& \textbf{97.84}
& \underline{37.36}
& \underline{56.79}
& \underline{36.73} \\
\textbf{Qwen-Audio-3.0-Gen-Preview}
& \underline{87.35}
& \textbf{43.12}
& \textbf{65.74}
& \textbf{43.83} \\

\bottomrule
\end{tabular*}
\end{table}

Both judges produce the same ranking pattern. Qwen-Audio-3.0-Gen-Preview
records mIoU values of 43.73 and 43.12 under Gemini-3.1-Pro-Preview and
Qwen3.5-Omni-Plus, respectively, compared with 38.48 and 37.36 for
Seed-Audio-1.0. The proposed model also records higher rates at both IoU
thresholds, whereas the baseline has higher event recall; this is a descriptive
coverage--localization trade-off, and no uncertainty estimates are reported.

\FloatBarrier

As the final task-level evaluation, we use SongBench~\cite{wu2026songbench} to
extend the analysis to long-range structure and rendering quality across seven
complementary components: melody, arrangement, musicality, vocal quality,
instrumental quality, mixing, and structure.

\begin{table}[H]
\caption{SongBench~\cite{wu2026songbench} component scores for base pre-trained
checkpoints evaluated on the same small evaluation set. The proposed model uses
roughly an order of magnitude less data for this benchmark family than the
in-house model. Parentheses give scores reported
in~\cite{dai2026fullsong} for the final in-house system after reward-based
post-training, including RL, on a substantially larger separate benchmark; these italic
values are not directly comparable and are excluded from the base-checkpoint
ranking. Bold and underlined values indicate the best and second-best
base-checkpoint scores in each component, respectively. Arr., Mus., Instr., and Struct. abbreviate
arrangement, musicality, instrumental quality, and structure.}
\label{tab:songbench}
\centering
\small
\setlength{\tabcolsep}{4.5pt}
\begin{tabularx}{\textwidth}{@{}>{\raggedright\arraybackslash}p{2.03in}*{7}{>{\centering\arraybackslash}X}@{}}
\toprule
\multirow{2}{*}{\textbf{Model}}
& \multicolumn{7}{c}{\textbf{SongBench} $\uparrow$} \\
\cmidrule(lr){2-8}
& \shortstack{\scriptsize\textbf{Melody}}
& {\scriptsize\textbf{Arr.}}
& {\scriptsize\textbf{Mus.}}
& \shortstack{\scriptsize\textbf{Vocal}}
& {\scriptsize\textbf{Instr.}}
& \shortstack{\scriptsize\textbf{Mixing}}
& {\scriptsize\textbf{Struct.}} \\
\midrule
\shortstack[l]{Dedicated in-house AR model\\[-1pt]\strut}
& \shortstack{\textbf{5.750}\\[-1pt]\textit{(7.200)}}
& \shortstack{\textbf{6.467}\\[-1pt]\textit{(7.388)}}
& \shortstack{\underline{4.548}\\[-1pt]\textit{(6.368)}}
& \shortstack{\textbf{5.208}\\[-1pt]\textit{(7.623)}}
& \shortstack{\underline{6.448}\\[-1pt]\textit{(7.352)}}
& \shortstack{\underline{5.820}\\[-1pt]\textit{(7.305)}}
& \shortstack{\textbf{5.425}\\[-1pt]\textit{(6.965)}} \\
\mbox{\textbf{Qwen-Audio-3.0-Gen-Preview}}
& \underline{5.608} & \underline{6.331} & \textbf{4.813}
& \underline{5.154} & \textbf{6.559} & \textbf{5.884}
& \underline{5.322} \\
\bottomrule
\end{tabularx}
\end{table}

The comparison uses the unified Qwen-Audio-3.0-Gen-Preview checkpoint and a
dedicated in-house AR baseline~\cite{dai2026fullsong}. With roughly an order of
magnitude less data for this benchmark family, the proposed model remains
numerically close across all seven components and leads in musicality,
instrumental quality, and mixing. The comparison is not controlled: the systems
differ in training data, representations, objectives, and model scope, and the
base-checkpoint evaluation uses a small sample with no uncertainty estimates.

\FloatBarrier

\subsection{VAE Component Evaluation}
\label{sec:vae-evaluation}

After the task-level evaluations above, we evaluate the VAE in
Section~\ref{sec:audio-vae} through waveform reconstruction and controlled
downstream probe. The probe trains a task-specific generator against different
target representations; it does not evaluate the complete
Qwen-Audio-3.0-Gen-Preview system used in the preceding subsections.

\paragraph{Evaluation setup.}
For reconstruction, we compare DAC~\cite{kumar2023dac},
EAR-VAE~\cite{wang2025earvae}, SAME-L~\cite{parker2026same}, Stable Audio
Open~\cite{evans2025sao}, Semantic-VAE~\cite{niu2025semanticvae},
LoSATok~\cite{zhang2026losatok}, and HoliTok~\cite{li2026holitok}. We evaluate
reconstruction on Seed-TTS EN and
ZH~\cite{a2024seedtts}, the Song Describer Dataset~\cite{manco2023sdd},
MuChin~\cite{wang2024muchin}, and AudioCaps~\cite{kim-etal-2019-audiocaps}. We
group representations by latent
frame rate into High-Rate ($\geq 40$\,Hz) and Low-Rate ($<40$\,Hz) systems. In
the reconstruction tables, SR and LR denote waveform sample rate and latent
frame rate.

For the downstream probe, we use an F5-style conditional flow-matching
generator~\cite{chen2025f5} on LibriSpeech-PC-200~\cite{panayotov2015librispeech}
and include LoSATok~\cite{zhang2026losatok} as a semantic latent baseline. Every
target representation uses the same
generator architecture, training data, and optimization budget. We denote the
reconstruction-only checkpoint as \emph{Qwen-Audio-Gen-VAE (Acoustic)} and its
semantically continued counterpart as \emph{Qwen-Audio-Gen-VAE}; Sem. indicates whether
the target representation uses semantic supervision.

\paragraph{Reconstruction quality.}
We use ViSQOL for perceptual quality; Mel and STFT distances for spectral
fidelity; SI-SDR for signal-level reconstruction; PESQ and STOI for speech
quality and intelligibility; and CCPC and ICPC for stereo phase coherence.

\newcolumntype{V}{>{\centering\arraybackslash}X}

\begin{table}[!htbp]
    \centering
    \caption{Reconstruction on Seed-TTS EN and ZH. Bold and underlined
    values indicate the best and second-best Low-Rate results in each metric,
    respectively.}
    \label{tab:vae-speech-reconstruction}
    \scriptsize
    \setlength{\tabcolsep}{3pt}
    \renewcommand{\arraystretch}{1.08}
    \begin{tabularx}{\textwidth}{>{\raggedright\arraybackslash}p{0.19\textwidth}*{8}{V}}
        \toprule
        \multirow{2}{*}{\textbf{Model}}
        & \multicolumn{8}{c}{\textbf{Seed-TTS EN}} \\
        \cmidrule(lr){2-9}
        & \textbf{SR} & \textbf{LR} & \textbf{ViSQOL} $\uparrow$ & \textbf{Mel} $\downarrow$ & \textbf{STFT} $\downarrow$
        & \textbf{SI-SDR} $\uparrow$ & \textbf{PESQ} $\uparrow$ & \textbf{STOI} $\uparrow$ \\
        \midrule
        \multicolumn{9}{l}{\textit{High-Rate}} \\
        DAC & 44.1\,kHz & 86\,Hz & 4.3919 & 0.4693 & 1.0233 & 11.2306 & 3.7712 & 0.9691 \\
        EAR-VAE & 44.1\,kHz & 43\,Hz & 4.2253 & 0.5532 & 1.0599 & 13.2584 & 3.4765 & 0.9660 \\
        SAME-L & 44.1\,kHz & 43\,Hz & 3.9208 & 0.6918 & 1.2487 & 13.4237 & 2.7375 & 0.9535 \\
        Semantic-VAE & 16\,kHz & 40\,Hz & 3.2171 & 0.7818 & 1.7031 & 9.6977 & 3.6097 & 0.9738 \\
        \midrule
        \multicolumn{9}{l}{\textit{Low-Rate}} \\
        HoliTok & 48\,kHz & 25\,Hz & \underline{4.3905} & \textbf{0.4248} & \textbf{1.0050} & 9.3901 & \textbf{3.9208} & \textbf{0.9783} \\
        Stable Audio Open & 44.1\,kHz & 21.5\,Hz & 4.2587 & 0.6587 & 1.1594 & 6.7421 & 2.5371 & 0.9264 \\
        Qwen-Audio-Gen-VAE (Acoustic) & 48\,kHz & 25\,Hz & 4.2282 & \underline{0.5683} & \underline{1.0669} & \underline{12.0236} & 3.6142 & 0.9687 \\
        Qwen-Audio-Gen-VAE & 48\,kHz & 25\,Hz & \textbf{4.3927} & 0.5788 & 1.0796 & \textbf{12.1295} & \underline{3.6455} & \underline{0.9708} \\
        \specialrule{0.8pt}{6pt}{2pt}
        \multirow{2}{*}{\textbf{Model}}
        & \multicolumn{8}{c}{\textbf{Seed-TTS ZH}} \\
        \cmidrule(lr){2-9}
        & \textbf{SR} & \textbf{LR} & \textbf{ViSQOL} $\uparrow$ & \textbf{Mel} $\downarrow$ & \textbf{STFT} $\downarrow$
        & \textbf{SI-SDR} $\uparrow$ & \textbf{PESQ} $\uparrow$ & \textbf{STOI} $\uparrow$ \\
        \midrule
        \multicolumn{9}{l}{\textit{High-Rate}} \\
        DAC & 44.1\,kHz & 86\,Hz & 4.3189 & 0.4370 & 0.9466 & 12.3867 & 3.8113 & 0.9654 \\
        EAR-VAE & 44.1\,kHz & 43\,Hz & 4.2215 & 0.4624 & 0.8864 & 14.8875 & 3.6955 & 0.9643 \\
        SAME-L & 44.1\,kHz & 43\,Hz & 3.8982 & 0.5460 & 0.9724 & 14.9420 & 3.0809 & 0.9488 \\
        Semantic-VAE & 16\,kHz & 40\,Hz & 2.9406 & 0.8306 & 1.8793 & 10.3511 & 3.6086 & 0.9695 \\
        \midrule
        \multicolumn{9}{l}{\textit{Low-Rate}} \\
        HoliTok & 48\,kHz & 25\,Hz & \underline{4.3436} & \textbf{0.3735} & \textbf{0.8591} & 10.7590 & \textbf{3.9951} & \textbf{0.9759} \\
        Stable Audio Open & 44.1\,kHz & 21.5\,Hz & 4.1953 & 0.6067 & 1.0325 & 8.4022 & 2.6192 & 0.9246 \\
        Qwen-Audio-Gen-VAE (Acoustic) & 48\,kHz & 25\,Hz & 4.2460 & \underline{0.4917} & \underline{0.9235} & \textbf{13.4252} & \underline{3.8554} & 0.9676 \\
        Qwen-Audio-Gen-VAE & 48\,kHz & 25\,Hz & \textbf{4.3535} & 0.5105 & 0.9274 & \underline{13.4169} & 3.8533 & \underline{0.9684} \\
        \bottomrule
    \end{tabularx}
\end{table}

\begin{table}[!htbp]
    \centering
    \caption{Reconstruction on the Song Describer Dataset and MuChin.
    Phase-coherence values are unavailable for the mono-native LoSATok. Bold
    and underlined values indicate the best and second-best Low-Rate results in
    each metric, respectively.}
    \label{tab:vae-structured-audio-reconstruction}
    \scriptsize
    \setlength{\tabcolsep}{3pt}
    \renewcommand{\arraystretch}{1.08}
    \begin{tabularx}{\textwidth}{>{\raggedright\arraybackslash}p{0.19\textwidth}*{8}{V}}
        \toprule
        \multirow{2}{*}{\textbf{Model}}
        & \multicolumn{8}{c}{\textbf{Song Describer Dataset}} \\
        \cmidrule(lr){2-9}
        & \textbf{SR} & \textbf{LR} & \textbf{ViSQOL} $\uparrow$ & \textbf{Mel} $\downarrow$ & \textbf{STFT} $\downarrow$
        & \textbf{SI-SDR} $\uparrow$ & \textbf{CCPC} $\uparrow$ & \textbf{ICPC} $\uparrow$ \\
        \midrule
        \multicolumn{9}{l}{\textit{High-Rate}} \\
        DAC & 44.1\,kHz & 86\,Hz & 4.1981 & 0.8718 & 1.3921 & 10.8695 & 0.7212 & 0.7256 \\
        EAR-VAE & 44.1\,kHz & 43\,Hz & 4.0576 & 0.4589 & 0.9725 & 11.9368 & 0.7909 & 0.7231 \\
        SAME-L & 44.1\,kHz & 43\,Hz & 4.0121 & 0.5038 & 1.1843 & 11.7872 & 0.7848 & 0.7151 \\
        \midrule
        \multicolumn{9}{l}{\textit{Low-Rate}} \\
        LoSATok & 16\,kHz & 25\,Hz & 2.2283 & 1.9373 & 3.4818 & -31.7331 & -- & -- \\
        Stable Audio Open & 44.1\,kHz & 21.5\,Hz & 4.0653 & 0.6009 & 1.2062 & 6.0569 & 0.7047 & 0.5708 \\
        Qwen-Audio-Gen-VAE (Acoustic) & 48\,kHz & 25\,Hz & \textbf{4.3010} & \textbf{0.4534} & \textbf{1.0752} & \textbf{11.4310} & \textbf{0.7900} & \textbf{0.6922} \\
        Qwen-Audio-Gen-VAE & 48\,kHz & 25\,Hz & \underline{4.2925} & \underline{0.5184} & \underline{1.1615} & \underline{10.8145} & \underline{0.7787} & \underline{0.6812} \\
        \specialrule{0.8pt}{6pt}{2pt}
        \multirow{2}{*}{\textbf{Model}}
        & \multicolumn{8}{c}{\textbf{MuChin}} \\
        \cmidrule(lr){2-9}
        & \textbf{SR} & \textbf{LR} & \textbf{ViSQOL} $\uparrow$ & \textbf{Mel} $\downarrow$ & \textbf{STFT} $\downarrow$
        & \textbf{SI-SDR} $\uparrow$ & \textbf{CCPC} $\uparrow$ & \textbf{ICPC} $\uparrow$ \\
        \midrule
        \multicolumn{9}{l}{\textit{High-Rate}} \\
        DAC & 44.1\,kHz & 86\,Hz & 4.1430 & 0.9624 & 1.6574 & 9.9117 & 0.6947 & 0.7033 \\
        EAR-VAE & 44.1\,kHz & 43\,Hz & 4.0065 & 0.5829 & 1.2510 & 10.7052 & 0.7821 & 0.7024 \\
        SAME-L & 44.1\,kHz & 43\,Hz & 4.0289 & 0.6971 & 1.7745 & 10.2689 & 0.7654 & 0.6880 \\
        \midrule
        \multicolumn{9}{l}{\textit{Low-Rate}} \\
        LoSATok & 16\,kHz & 25\,Hz & 2.2051 & 1.6715 & 2.8214 & -32.1244 & -- & -- \\
        Stable Audio Open & 44.1\,kHz & 21.5\,Hz & 4.0009 & 0.6942 & 1.4057 & 5.1473 & 0.6905 & 0.5428 \\
        Qwen-Audio-Gen-VAE (Acoustic) & 48\,kHz & 25\,Hz & \underline{4.2828} & \textbf{0.4778} & \textbf{1.1315} & \textbf{10.2356} & \textbf{0.7785} & \textbf{0.6729} \\
        Qwen-Audio-Gen-VAE & 48\,kHz & 25\,Hz & \textbf{4.3064} & \underline{0.5335} & \underline{1.2268} & \underline{9.9425} & \underline{0.7734} & \underline{0.6641} \\
        \bottomrule
    \end{tabularx}
\end{table}

\begin{table}[!htbp]
    \centering
    \caption{Reconstruction on AudioCaps. Bold and underlined
    values indicate the best and second-best Low-Rate results in each metric,
    respectively.}
    \label{tab:vae-audio-reconstruction}
    \scriptsize
    \setlength{\tabcolsep}{3pt}
    \renewcommand{\arraystretch}{1.08}
    \begin{tabularx}{0.88\textwidth}{>{\raggedright\arraybackslash}p{0.22\textwidth}*{6}{V}}
        \toprule
        \multirow{2}{*}{\textbf{Model}}
        & \multicolumn{6}{c}{\textbf{AudioCaps}} \\
        \cmidrule(lr){2-7}
        & \textbf{SR} & \textbf{LR} & \textbf{ViSQOL} $\uparrow$ & \textbf{Mel} $\downarrow$ & \textbf{STFT} $\downarrow$ & \textbf{SI-SDR} $\uparrow$ \\
        \midrule
        \multicolumn{7}{l}{\textit{High-Rate}} \\
        DAC & 44.1\,kHz & 86\,Hz & 4.2854 & 0.5346 & 1.1450 & 5.9002 \\
        EAR-VAE & 44.1\,kHz & 43\,Hz & 4.1789 & 0.6078 & 1.1402 & 7.2334 \\
        SAME-L & 44.1\,kHz & 43\,Hz & 3.9516 & 0.7012 & 1.3015 & 7.6653 \\
        \midrule
        \multicolumn{7}{l}{\textit{Low-Rate}} \\
        LoSATok & 16\,kHz & 25\,Hz & 3.1208 & 1.2495 & 2.5070 & -34.7308 \\
        Stable Audio Open & 44.1\,kHz & 21.5\,Hz & 4.1664 & 0.6731 & \underline{1.1796} & 0.4401 \\
        Qwen-Audio-Gen-VAE (Acoustic) & 48\,kHz & 25\,Hz & \underline{4.2538} & \textbf{0.6075} & \textbf{1.1708} & \textbf{6.5950} \\
        Qwen-Audio-Gen-VAE & 48\,kHz & 25\,Hz & \textbf{4.3450} & \underline{0.6411} & 1.1849 & \underline{6.4642} \\
        \bottomrule
    \end{tabularx}
\end{table}

\begin{samepage}

Qwen-Audio-Gen-VAE (Acoustic) achieves the best Low-Rate Mel and STFT distances
on the Song Describer Dataset, MuChin, and AudioCaps, while remaining
competitive on Seed-TTS EN and ZH despite its 25\,Hz latent rate. HoliTok
performs better on several Seed-TTS metrics.

Semantic continuation improves ViSQOL on Seed-TTS EN and ZH, MuChin, and
AudioCaps, while leaving the Song Describer Dataset result nearly unchanged. It
introduces modest degradation in spectral and stereo-coherence metrics, but
largely preserves the overall reconstruction quality.

\paragraph{Controlled downstream probe.}
On LibriSpeech-PC-200, we measure WER, speaker similarity (SIM), and UTMOS
after hundreds of thousands of training steps.
\end{samepage}

\begin{table}[!htbp]
    \centering
    \caption{Controlled generation probe on LibriSpeech-PC-200 after hundreds
    of thousands of training steps. Bold and underlined values indicate the best and
    second-best results in each column, respectively.}
    \label{tab:vae-tts-generation}
    \scriptsize
    \setlength{\tabcolsep}{8pt}
    \renewcommand{\arraystretch}{1.08}
    \resizebox{0.72\textwidth}{!}{%
    \begin{tabular}{lcccc}
        \toprule
        \textbf{Target representation} & \textbf{Sem.} & \textbf{WER (\%)} $\downarrow$ & \textbf{SIM} $\uparrow$ & \textbf{UTMOS} $\uparrow$ \\
        \midrule
        Mel spectrogram & N & 19.69 & 0.466 & 1.769 \\
        Stable Audio Open & N & 12.33 & 0.504 & 1.892 \\
        Qwen-Audio-Gen-VAE (Acoustic) & N & 7.71 & 0.507 & 3.015 \\
        \midrule
        Semantic-VAE & Y & \textbf{2.96} & \underline{0.594} & 3.034 \\
        LoSATok & Y & \underline{3.30} & \textbf{0.622} & \textbf{3.411} \\
        Qwen-Audio-Gen-VAE & Y & 4.08 & 0.488 & \underline{3.367} \\
        \bottomrule
    \end{tabular}%
    }
\end{table}

\FloatBarrier

Qwen-Audio-Gen-VAE (Acoustic) records better WER, SIM, and UTMOS than both the Mel
spectrogram and Stable Audio Open targets.
Relative to Qwen-Audio-Gen-VAE (Acoustic), semantic continuation lowers WER and
raises UTMOS but reduces SIM; Semantic-VAE and LoSATok have lower WER and higher
SIM than Qwen-Audio-Gen-VAE. Because the generator architecture, training data,
and optimization budget are held fixed, these differences concern target
representations rather than the proposed model as a whole.

\section{Conclusion}
Qwen-Audio-3.0-Gen-Preview unifies heterogeneous standalone and mixed-scene
audio through a single non-autoregressive generation path. Its shared textual
interface conditions a DiT in the continuous latent space of a common VAE, whose
48\,kHz stereo representation operates at 25\,Hz and is augmented by semantic
continuation after reconstruction training. The VAE decodes the generated
latent sequence directly into the complete mixed waveform. Real annotations,
synthetic recipes, and prompt-enhanced requests use compatible structured
records rendered as textual conditions, while semantic views, role-bundle
integrity, and classifier-free guidance expose complementary information
without breaking source--dialogue relations.

The proposed model shows strong speaker similarity on the public
reference-conditioned benchmark. Its clearest comparative strengths on the
multi-speaker and rich-timeline benchmarks are cross-turn consistency and
temporal localization, respectively. On AudioCaps, its clearest advantages occur
under LALM and AudioBox assessments. Reconstruction measurements and the
LibriSpeech-PC-200 controlled probe support the utility of the low-frame-rate VAE
representation. Together, these results highlight the potential for temporally
controllable audio generation without task-specific branches.

\section*{Acknowledgements}

We thank Zhifu Gao, Yiping Peng, Shiming Wang, Yang Xiang, Jianwei Yu, Jixing Yu,
Peiyun Zeng, and Nan Zhao for their contributions to data curation, system
development, evaluation, and infrastructure support.

\textit{(Names are listed alphabetically by family name.)}

{\small
\bibliographystyle{plain}
\bibliography{reference}
}

\end{document}